\documentclass[12pt,preprint]{mnras}
\usepackage{amsmath}

\usepackage{graphicx}	
\usepackage{amssymb}	

\title[Formation of symbiotic X-ray binaries via AIC of white dwarfs in binaries]
{A promising formation channel for symbiotic X-ray binaries: cases of IGR J17329-2731 and 4U 1700+24}

\author[I. Ablimit]
{Iminhaji Ablimit$^{1,2}$\thanks{E-mail: \href{mailto:iminhaji@nao.cas.cn}{iminhaji@nao.cas.cn}}
\\
$^{1}$Key Laboratory for Optical Astronomy, National Astronomical Observatories,
Chinese Academy of Sciences, Beijing 100012, China\\
$^{2}$Department of Astronomy, Kyoto University, Kitashirakawa-Oiwake-cho, Sakyo-ku, Kyoto 606-8502, Japan}

\begin{document}
\label{firstpage}
\pagerange{\pageref{firstpage}--\pageref{lastpage}}
\maketitle

\begin{abstract}

Recent observations demonstrate that the symbiotic X-ray binary (SyXB) IGR J17329-2731 contains a 
highly magnetized neutron star (NS) which accretes matter through the wind from its 
giant star companion, and suggest that 4U 1700+24 may also have a highly magnetized NS.
Accretion-induced collapse (AIC) from oxygen-neon-magnesium white dwarf (ONeMg WD) + red giant (RG) star binaries
is one promising channel to form these SyXBs, while other long standing
formation channels have difficulties to produce these SyXBs.
By considering non-magnetic and magnetic ONeMg WDs, I investigate the evolution of ONeMg WD + RG binaries
with the \textsc{mesa} stellar evolution code for producing SyXBs with non-magnetic or magnetized NSs.
In the pre-AIC evolution with magnetic confinement, the mass accumulation
efficiency of the accreting WD is increased at low mass transfer rate compared to the non-magnetic case.
The newborn NSs formed via AIC of highly magnetized WDs could inherit the
large magnetic field through conservation of magnetic flux,
and the systems could have a long age compatible
with that of the red giant companions.
These young and highly magnetized NSs could accrete matters from the stellar
wind of the giant companions to that shine as those observed SyXBs, and could preserve
their high magnetic field during this time. The \textsc{mesa} calculation results show that the initial parameter (initial RG mass and orbital
period) space for the AIC with magnetic confinement to form SyXBs with highly magnetized NSs
shifts to be lower and narrower compared to that of the no magnetic confinement case. 

\end{abstract}




\section{Introduction}

Symbiotic X-ray binaries (SyXBs) are identified as a rare class of low-mass
X-ray binaries (LMXBs), and mainly composed of red giant (RG) stars and neutron stars (NSs)
(e.g., Masetti et al. 2006; Kaplan et al. 2007;
Corbet et al. 2008; Nespoli et al. 2008; Bozzo et al. 2011, 2012, 2013). Recently,
two known SyXBs have been reported and suggested to have highly magnetized NSs: (1) IGR J17329-2731.
Bozzo et al. (2018) analyzed the data collected quasi-simultaneously with
XMM-Newton and NuSTAR, and they estimated, according to the observational results, that this SyXB contains
an NS with a magnetic field strength of $\sim 2.4\times10^{12}$ Gauss. (2) 4U 1700$+$24.
Bozzo et al. (2022) performed broad-band X-ray and soft $\gamma$-ray spectroscopy of 4U 1700$+$24,
and found the presence of cyclotron scattering features. Based on their observational data analysis,
they suggest that 4U 1700$+$24 may have a highly magnetized NS with a magnetic field strength of $\sim 1.4\times10^{12}$ Gauss.
There is a very interesting and unclear question: how are these highly magnetized NS SyXBs as members of LMXBs are formed?

NSs in LMXBs can be formed via core-collapse supernovae during binary stellar evolution,
in this case NSs usually have low-mass main-sequence (MS) donors ($<1 M_\odot$).  In addition, LMXBs were suggested as the product of
the common envelope phase caused by unstable mass transfer from high-mass (donor's mass is larger than 10 $M_\odot$) X-ray
binaries (van den Heuvel 1994).
Then, it was suggested that LMXBs may have evolved from intermediate-mass (donor's mass is
between 3 and 7 $M_\odot$) X-ray binaries (King \& Ritter 1999 ; Podsiadlowski \& Rappaport 2000 ; Kolb et al. 2000).
In these formation routes, the donor stars fill their
Roche-lobe due to nuclear evolution or orbital angular
momentum loss, and an accretion disk may formed nearby around the NSs.
Spin periods of the NSs can be accelerated, and the magnetic field
of the NSs may experience decay by the mass accretion.
The magnetic field decay induced by the mass accretion is not well studied in detail,
however, it is generally believed that the amount of field decay is closely
related to the accreted mass (Taam \& van den Heuvel 1986, Shibazaki et al.
1989). As one class of LMXBs, the formation of SyXBs with strongly magnetized
NSs through the formation routes introduced above is difficult,
because it is hard to explain the presence of a highly magnetized NS in SyXB, and the main reasons are as follows:
1. In those formation routes, the mass transfer and accretion phase cannot be avoided when becoming an X-ray binary.
If a highly magnetized NS was born in LMXBs, the low mass MS star in this LMXB at
least needs 9-10 Gyr to become an RG and fill its Roche-lobe to start the mass transfer.
Then, the NS at least accretes some amount of the transferred mass from the donor and emits high energy particles,
and this accretion could decay the magnetic field of the NS.
2. If the newborn NS in LMXBs has both a strong magnetic field and very high rotational velocity,
the propeller effect may take place, and the NS ejects all matters transferred from the donor.
Thus, no accretion occurs in the case of the propeller effect, and the NS may retain its strong magnetic
field until the donor evolves to an RG.
However, there is also no high energy emissions in this case, and the binary cannot be an X-ray binary.

NSs also can form from WD binaries: (1) Merger-induced collapse of two WDs in a binary (Nomoto \& Iben 1985).
(2) Core-merger-induced collapse during the common envelope evolution of the WD binary (Ablimit et al. 2022).
(3) Accretion induced collapse (AIC) of the WD in a binary (e.g., Nomoto et al. 1979; Taam \& van den Heuvel 1986).
AIC of a highly magnetized WD in a binary may be
the promising solution for the formation puzzle of SyXBs with strongly magnetized
NSs.  AIC from oxygen/neon/magnesium (ONeMg) composition WD binaries has been
studied as one of the comprehensive formation pathways for some peculiar NS systems
including millisecond pulsars (e.g., Chanmugam \& Brecher 1987; Michel 1987;
Kulkarni \& Narayan 1988; Bhattacharya
\& van den Heuvel 1991; Yoon \& Langer 2005; Hurley et al. 2010; Sabach \& Soker 2014;
Tauris et al. 2013; Freire \& Tauris 2014; Ablimit \& Li 2015; Wang \& Liu 2020; Perets 2022),
redbacks/black widows (e.g., Ablimit 2019), and magnetars (Ablimit 2022).
If the WD in the AIC process is highly magnetized, the high magnetic
flux could be conserved and even could be amplified during the
AIC process (e.g., Duncan \& Thompson 1992).

The magnetic WDs have been observed in cataclysmic variables (CVs;
e.g., Ferrario et al. 2015), symbiotic binaries and super-soft X-ray
sources (Kahabka 1995; Sokoloski \& Bildsten 1999; Osborne
et al. 2001). However, it is worth noting that the AIC process
(including conservation of magnetic flux) and
evolution of highly magnetized WD binaries as the progenitors
of AIC or type Ia supernovae (SNe Ia) are not well-known from the aspect of both observation and theory. In addition, the origin of WDs' magnetic fields also remains unclear (see Mukhopadhyay \& Bhattacharya [2022] for a recent review on magnetized compact stars). Magnetic WDs could be the fossil remnants of their progenitors (e.g., Angel et al. 1981), they would be formed from the field amplification during the huliem-burning phase (Levy \& Rose 1974), or they could be generated from the binary interactions (Tout et al 2004; Nordhaus et al. 2011; Garcia-Berro et al. 2012). It has been recently suggested that single WDs' magnetic fields can be generated by a dynamo of crystallization in a process analogous to the Earth (Isern et al. 2017; Belloni et al. 2021; Schreiber et al. 2021; Camisassa et al. 2022; Ginzburg et al. 2022).

Although polars (CVs containing highly magnetized WDs) known to date have companion
stars with relatively lower masses which cannot be progenitors of AIC,
very recent and interesting observational results demonstrated that
highly magnetized WDs like polars can have a two-pole accretion mode compared
to non-magnetic or weakly magnetized WDs (Schwope
et al. 2021). Ablimit (2019; see also Ablimit \& Maeda 2019a,b) and Ablimit (2022) explored
the role of magnetism in hydrogen-rich or helium rich material accreting WD binaries
as progenitors of AIC (or SNe Ia), and they found that the initial donor mass - orbital
period space for producing AIC (or SNe Ia) can be different if magnetic confinement is achieved.

The main goal of the work is that I study the detailed pre-AIC evolution
of highly magnetized, intermediate and non-magnetic ONeMg WDs with RG companions as
a promising formation channel for SyXBs with highly magnetized NSs or with non-magnetic NSs.
By employing the \textsc{mesa} stellar evolution code, I simulate evolution of the ONeMg WD $+$ RG
binaries for cases with and without magnetic confinement.
The details for the one dimensional \textsc{mesa}
calculations of binary evolution including magnetic confinement model are described in Section 2.
In Section 3, I give and discuss our results. The conclusion is presented in Section 4.


\section{The evolution of WD $+$ red giant star binaries (pre-AIC) without and with the magnetic confinement}
\label{sec:model}

With the star and binary packages of version 15140 of \textsc{mesa} stellar evolution
code (Paxton et al. 2011, 2015), numerical simulations for the
detailed evolution of WD $+$ RG binaries are performed in this work.
First of all, in order to simulate the RG star models I employ the star package with
a typical Pop I composition with hydrogen abundance $X = 0.70$, He abundance
$Y = 0.28$ and metallicity $Z = 0.02$. In the the stellar evolution,
$initial\_zfracs = 6$ and $kappa\_file\_prefix =$ `a09' are set to
call the opacity tables which are built using the more recently available solar composition,
and the Henyey theory of convection with $mixing\_length\_alpha=1.8$ is used.
The stellar evolution is begun from a pre-main sequence model,
and the final model is saved to start the next step.
The evolution is continued
by loading the previously save model, and stops it when the central helium fraction $\geq 0.98$
(it has a helium core and RG radius at this stage).
RG star models with masses ($M_{\rm RG}$) ranging from 0.5 to 2.0 $M_\odot$ are built by using a mass interval of 0.1 $M_\odot$.

I use the binary package of \textsc{mesa} to simulate ONeMg WD $+$ RG binary evolution with
the initial orbital periods ($P_{\rm orb,i}$) of 0.5 - 2000 days.
With RG star models introduced above, and WDs are
simulated as a point mass with initial masses of 1.2 $M_\odot$.
Because giant stars have lower surface
gravities and extended atmospheres, it is relatively complicated when
estimating the mass-transfer rate via the Roche-lobe overflow (RLOF).
In this work, the computational method in Kolb and Ritter (1990) is adopted for
estimating the RLOF mass transfer rate for RG stars, which should generate reasonable results for
stars at different evolutionary stages,
\begin{multline}
{\dot{M}_{\rm RL}} =
-\dot{M}_{0}-2{\pi}F(q_2)\frac{R^3_{\rm RL}}{GM_{\rm RG}}\\
\times{\int^{P_{\rm RL}}_{P_{\rm ph}} {{\Gamma _1}^{1/2}{(\frac{2}{\Gamma_1 +1})}^{(\frac{\Gamma_1 +1}{2\Gamma_1 -2})}
{(\frac{\kappa_{\rm B} T}{m_{\rm p} \mu})}}\,\mathrm{d}P} ,
\end{multline}
where $\dot{M_{\rm RL}}$ is the RLOF mass transfer rate, $\Gamma _1$ is the first adiabatic exponent, and
$P_{\rm ph}$ and $P_{\rm RL}$ are the pressures at the photosphere and at the radius
when the radius of the donor is equal to its RLOF radius ($R_{\rm RL}$), respectively.
$T$ is the temperature of the donor,  $\kappa_{\rm B}$ is Boltzmann constant
and $\mu_{\rm ph}$ is the mean molecular weight.
$\dot{M}_{0}$ is
\begin{equation}
\dot{M}_{0} = \frac{2\pi}{\rm exp(1/2)}F(q_2)\frac{R^3_{\rm RL}}{GM_{\rm RG}}{(\frac{\kappa_{\rm B} T_{\rm eff}}{m_{\rm p} \mu_{\rm ph}})^{3/2}}\rho_{\rm ph},
\end{equation}
where $m_{\rm p}$ is the proton mass, and $T_{\rm eff}$ is the effective temperature of
the donor. $\mu_{\rm ph}$ and $\rho_{\rm ph}$ are the mean molecular weight and density at its photosphere.
For the fitting function $F({q_2})$ ($q_2 = M_{\rm accretor}/M_{\rm donor}$) and other physical parameters in the evolutions are
same as typical ones given in the instrumental papers of \textsc{mesa} (e.g., Paxton et al. 2015).
Stars in the giant phase have relatively higher wind mass loss compared to stars at the early stages; the prescription
in Reimers (1975) is used for calculating wind mass loss rate,
\begin{equation}
\dot{M}_{\rm{wind}} = 4\times10^{-13} \frac{L_{\rm RG}R_{\rm RG}}{M_{\rm RG}} {\eta_{\rm R}},
\end{equation}
where $L_{\rm RG}$ and $R_{\rm RG}$ are the luminosity and radius of the RG star,
and $\eta_{\rm R}=0.5$ is a typical value in the MESA code. The Reimers scheme and
Blocker scheme in \textsc{mesa} are set for the wind mass loss in the
giant phase by considering possible switch between RG and AGB branches (Paxton et al. 2015).

In the accretion phase of the WD binary, the crucial role of mass transfer rate
is that it decides whether the material transferred from the donor onto
the WD can stably burn to let the WD grow in mass. When the mass transfer rate (${\dot{M}_{\rm RL}}$) is
suitable to realize stable hydrogen and helium burning on the WD,
 the mass growth rate of the WD is,
\begin{equation}
\dot{M}_{\rm{WD}} = \eta_{\rm H} \eta_{\rm He} {\dot{M}_{\rm RL}}.
\end{equation}
Hillman et al. (2015, 2016) studied the efficiency of
hydrogen burning ($\eta_{\rm H}$) in a detailed way by considering the WD's properties, and I adopted their prescription.
Zero and non-zero values of $\eta_{\rm H}$ can be derived in this physical process: $\eta_{\rm H}=0.0$ (no mass growth) when
${\dot{M}_{\rm RL}} < \sim 5\times10^{-8}\,\rm{M_\odot\,yr^{-1}}$ and nova bursts occurs,
$0.0<\eta_{\rm H}\leq1.0$ and WD can grow in mass if $\sim 5\times10^{-8}\,\rm{M_\odot\,yr^{-1}}
\leq {\dot{M}_{\rm RL}} < 10^{-4}\,\rm{M_\odot\,yr^{-1}}$ (see Hillman et al. (2016) for more details),
and the binary enters the CE phase ($\eta_{\rm H}=0.0$) if ${\dot{M}_{\rm RL}} > 10^{-4}\,\rm{M_\odot\,yr^{-1}}$. Some of transferred mass can burn stably and some of them can be ejected from the WD if the mass transfer rate exceeds $\sim 10^{-6}\,M_\odot\,{\rm yr^{-1}}$ (see also Nomoto et al. 2007; Wolf et al. 2013).
The calculation methods of
Kato \& Hachisu (2004) for the mass accumulation efficiency of
helium ($\eta_{\rm He}$) are utilized in this work.
These mass retention efficiencies are very important,
and the prescriptions used in this work are widely recognized in the field.
The different episodes for carbon burning cannot significantly affect the growth of the WD (Brooks et al. 2017),
thus the carbon burning model does not have influence on our results.
Some transferred matter may be lost from the system as the optically thick wind when $0.0<\eta_{\rm H}<1.0$,
and it takes the angular momentum away with it. The angular momentum evolution caused by
all mass loss and other possible mechanisms (i.e. gravitational wave, magnetic braking) are included in the calculations (see Paxton et al. 2015).


Compared to the spherically symmetric accretion, a strongly magnetized WD could have the stream/confined
accretion and unique emissions (e.g., Fabian et al. 1977; Livio 1983; King \& Shaviv 1984; Hameury et al. 1986;
King 1993; Wickramasinghe \& Ferrario 2000; Wickramasinghe 2014; ; Ferrario et al. 2015;
Ablimit 2019). The accretion disc around the WD may be fully or partially disrupted
by the WD's strong dipolar magnetic field (e.g., Cropper 1990; Frank et al. 2002), and matter
transferred via the RLOF of the donor star follows magnetic field lines and falls down
onto the magnetic poles of the WD as an accretion column (see schematic figure of Ablimit (2019)).
In this case, the motion of the accreting matter could be dominated by the strong magnetic field of the WD,
 because the magnetic
pressure ($P_{\rm B}$) increases rapidly as the accreted material approaches
the WD's surface, and the magnetic pressure could exceed
the ram pressure ($P_{\rm ram}$) of the accreting matter at the magnetospheric radius ($R_{\rm M}; $Frank et al. 2002),
\begin{equation}
P_{\rm{B}} = \frac{(B R^3_{\rm WD})^2}{8\pi R^6_{\rm M}} \geq P_{\rm{ram}} = \frac{(2GM_{\rm WD})^{1/2}{\dot{M}_{\rm RL}}}{4\pi R^{5/2}_{\rm M}}.
\end{equation}
Thus, a minimal condition to have the magnetic confinement in the accretion process of the hydrogen-rich
material (Livio 1983) is that the magnetic field strength ($B$) of the WD satisfies,
\begin{multline}
B \geq 9.3\times10^{7} (\frac{R_{\rm WD}}{5\times10^8\,\rm cm})(\frac{P_{\rm b}}{5\times10^{19}\,\rm{dyne\,cm^{-2}}})^{7/10}\\
(\frac{M_{\rm WD}}{\rm{M_\odot}})^{-1/2}(\frac{\dot{M}_{\rm RL}}{10^{-10}\,\rm{M_\odot\,yr^{-1}}})^{-1/2}\, \rm{Gauss},
\end{multline}
where the pressure at the base of accreted matter is taken as $P_{\rm b} = 5\times10^{19}\,\rm{dyne\,cm^{-2}}$ (Livio 1983);
the formula in Nauenberg (1972) is adopted for the mass ($M_{\rm WD}$) -- radius ($R_{\rm{WD}}$) relation of the WD.

Once the magnetic field strength of WD meets the minimal condition, the nova
burst can be suppressed and the magnetic confinement can be realized in the
accretion phase of a sufficiently magnetized WD even at a low mass transfer rate.
Then, the highly magnetized WD can have the isotropic pole-mass transfer rate ($\dot{M}_{\rm{p}}$),
which is (see also Ablimit \& Maeda 2019a),
\begin{equation}
\dot{M}_{\rm{p}} =  \frac{S}{\Delta \rm{S}} {\dot{M}_{\rm RL}},
\end{equation}
where the surface area of a WD is $S=4\pi R^2_{\rm WD}$, and the area of the polar caps (on which material is accreted) can be estimated as $\Delta {S}=2\pi R^2_{\rm WD}\,\rm{sin^2{\beta}}$ ($\beta$ is the half-angle of the polar cap; Frank et al. 2002). If the relation between two angles ($\beta$ and $\theta$ which is the angle between the rotation axis and the magnetic field axis; see  Frank et al. (2002) for more details) is used, then the ratio is ${S}/{\Delta {S}}={{2 R_{\rm m}}}/({R_{\rm WD}\,\rm{cos^2{\theta}})}$, where I take $\theta = 0$ (see also Ablimit \& Maeda (2019a), Ablimit (2019) and Ablimit (2022) for more details) in the calculations.
${R_{\rm m}}$ is the magnetospheric radius which is related with Alfven radius ${R_{\rm A}}$ (Lamb et al.\ 1973; Norton \& Watson 1989; Frank et al.\ 2002),
\begin{equation}
R_{\rm m}= \phi {R_{\rm A}} = \phi\, 2.7\times10^{10} {M^{-1/7}_{\rm WD}} {\dot{M}^{-2/7}_{\rm acc}} {{\mu}^{4/7}}\,\rm{cm},
\end{equation}
where $\phi$ is a parameter ($\leq 1$) that takes into account the departure from the spherically symmetric case, ${\mu}= {B} {R^3_{\rm WD}}$ is the magnetic moment of the WD
in units of $10^{33}$ G $\rm{cm}^3$, $\dot{M}_{\rm acc}$ is the mass-accretion rate (which is the actual mass transfer rate $\dot{M}_{\rm RL}$ in this case) in units of $10^{16}$ g $\rm{s}^{-1}$ and $M_{\rm WD}$ is the WD mass in solar units.
$\dot{M}_{\rm{p}}$ is the rate to calculate the hydrogen burning efficiency in the magnetic confinement model,
while $\dot{M}_{\rm RL}$ is one to derive the hydrogen burning efficiency in the case of no magnetic confinement.

In this work, it is assumed that the ONeMg WD will collapse into an NS when it grows
in mass and reaches the Chandrasekhar limit mass of 1.38 $M_\odot$.
This limit mass to collapse may be higher for relatively fast rotating ONeMg WDs but not all WDs rotate that fast
(Yoon \& Langer 2005; Freire \& Tauris 2014; Ablimit \& Li 2015), and it will not affect the main goal of this work. It is also worth noting that the magnetic field generation for some WDs may need the fast rotation (e.g., Ginzburg et al. 2022).
The rotation of WDs is not studied in the work, and the main focus is to examine the formation of SyXBs with
highly magnetized NSs with the assumption of magnetic flux conservation during the AIC of highly magnetized WDs.
In the binary evolution, the fixed values for non-, intermediate-
and high-magnetic field strengths of the WDs are taken
as 0, $1\times10^4$ G and  $2.5\times10^7$ G \text{(e.g., Livio 1983; Ferrario et al. 2015)}, respectively.


\section{Results and discussion}

\subsection{Pre-AIC evolution}

Figure 1 shows initial parameter spaces of WD$+$RG binary systems, the
upper panel is for systems having WDs with non- (0) or intermediate
($1\times10^4$ G) magnetic field strengths (no magnetic confinement),
and the lower panel is for the highly magnetized ($2.5\times10^7$ G) WD binaries (magnetic confinement case).
In the initial orbital period ($P_{\rm orb,i}$)-secondary mass ($M_{\rm RG, i}$) planes, red star symbols mark the binaries which lead to the AIC of ONeMg WDs,
while green circles, black squares and crosses represent binaries that cannot produce
AIC due to inefficient mass transfer or initiation of nova bursts.
The WD binaries with higher mass RG stars and too high mass transfer rate may undergo the CE evolution,
and this kind of system is not marked in the figure.
The ranges of initial donor mass and orbital
period for the no magnetic confinement case are
$1.1-1.8\,M_\odot$ and $10-750$ days respectively.
For the magnetic confinement case,
they are $0.7-1.8\,M_\odot$ and $5-120$ days. The initial parameter space for the AIC
shifts to lower regions compared to the no magnetic confinement case, the one main reason is that the mass loss from the system is generally higher for the higher values of $\dot{M}_{\rm p}$ (the WD eject some of the transferred mass when $\dot{M}_{\rm p}> 10^{-6}\,M_\odot\,{\rm yr^{-1}}$).
Compared with the previous work (e.g., Tauris et al. 2013;
Wang et al. 2022), the progenitors of AIC in this work have
relatively shorter initial orbital periods
and a wider range of initial masses of the RG stars.
The main reason causing the different results is that the calculation method for RLOF mass transfer rate is different.
Besides, I considered both the wind mass loss and RLOF mass transfer in the evolution while
other works only considered one of them in their works. The RG in the pre-AIC evolution can lose a certain amount of mass through the
stellar wind before the RLOF, and this mass loss could help the system have a relatively stable RLOF mass transfer.
In the pre-AIC evolution, it is also worth noting that the wind mass loss rate before the RLOF is less than
$3\times10^{-8}\,\rm{M_\odot\,yr^{-1}}$ most of the time (besides only a small fraction of the mass lost
from the spherically (isotropic) stellar wind moves toward the WD), thus the RLOF mass transfer
rate dominates the mass accretion phase in the WD binary evolution.

Detailed evolution of a WD + RG binary in a 100 day orbit is shown in Figure 2,
and initial masses of the WD and RG star are 1.0 $M_\odot$ and $M_{\rm WD,i}=1.2 M_\odot$,
respectively. The RG star loses mass through wind during the evolution,
and the RG star expands and overflows its Roche lobe while burning the helium in its core.
Without magnetic confinement in this example (left panels of Figure 3),
the accreted hydrogen-rich material burns unstably on the WD surface and nova outbursts happen,
thus the nova bursts prevent the mass growth of the WD by blowing away the material transferred onto the WD.
As a result of this mass loss and the RG wind mass loss which affect
the angular momentum evolution of the system, the orbit of the binary is wider.
The WD's mass stays still without magnetic confinement, and the orbital period evolves from 100 to 920
days due to the angular momentum taken by these mass losses away from the binary.
Under the effect of WD's strong magnetic field (right panels of Figure 3),
the transferred matter can be confined in the polar cap regions, and the
higher polar mass transfer rate (red line) alters the mass accretion phase of the binary.
The main difference of these two cases with
the same RLOF mass transfer from the donor is that the burning efficiency of accreted hydrogen-rich
material per unit area on the WD can be higher than that in the spherical accretion (non-magnetic) case.
Therefore, the WD is gaining mass smoothly as the RG star is losing its mass under the the magnetic
confinement case, and it finally reaches the mass limit to explode as an SN Ia
(right panels of Figure 3). Correspondingly, the orbital period evolves in a much slower way,
because the system lost only a small amount of mass through the RG stellar
wind, and the binary with the magnetic confinement evolves further until the SN Ia occurs, so that
the radius of the RG star expands more than that of the RG star in the no magnetic confinement case.
Evolutionary outcomes are very different in this example due to the magnetic field of the WD.

In Figure 3, I show another example of WD $+$ RG binary evolution
which the WD ($M_{\rm WD,i}=1.2 M_\odot$) can reach $M_{\rm Ch}$ with
and without the magnetic confinement cases. The mass transfer in both cases proceeds on a
thermal timescale, the ratio of the mass of the RG to that of the WD is lower than the critical mass ratio
when the RG fills its Roche lobe, and the binary
systems undergo stable mass transfer until the WDs collapse.
In this binary, the wind mass loss rate of an RG star with an initial
mass of 1.3 $M_\odot$ is higher than
$10^{-10}\,\rm{M_\odot\,yr^{-1}}$ and up to $10^{-8}\,\rm{M_\odot\,yr^{-1}}$;
it takes more angular momentum away from the binary, thus the wind
mass loss widens the orbit significantly compared to that of the previous example.
With the higher mass donor (higher
RLOF mass transfer rate). The polar accretion rate is also much
higher in the magnetic confinement case (red line), and this higher polar accretion rate limits the steady
nuclear burning on the WD's polar cap regions (this means the magnetic field cannot
efficiently/sufficiently confine all transferred material in the polar cap regions). Thus, some of the transferred
 mass via RLOF would be lost from the binary and taken away the angular momentum with it
due to the higher polar accretion rate (red line in right panels; the mass loss rate will be higher if $\dot{M}_{\rm p}$ is higher [when it exceeds $10^{-6}\,M_\odot\,{\rm yr^{-1}}$], and some of the transferred matter can not burn stably on the WD and escape from the WD.), and the RG donor star
loses more mass in order to let the WD grow in mass to $M_{\rm Ch}$.
This mass loss also widen the binary orbit more compared to that of the no magnetic
confinement case (left panels), and the surviving companion star in
the magnetic confinement case is less massive than in the non-magnetic model.
The different mass accretion modes make the binary evolution run in different ways, so that
 properties of surviving companions are also different.

\subsection{Post-AIC evolution: Formation of SyXBs with highly magnetized NSs}

When ${M}_{\rm WD} = M_{\rm Ch}$, I assume that the WD collapses to be an NS with a gravitational mass of
1.25 $M_\odot$ (e.g., Schwab et al. 2010).
The orbital separation becomes wider due to the sudden mass-loss. According to the prescription
of angular momentum conservation, the relationship
between the orbital separations just before ($ a_0$) and after ($ a$) the collapse (Verbunt et al. 1990) is calculated as,
\begin{equation}
\frac{a}{a_0} =  \frac{M_{\rm WD} + M_2}{M_{\rm NS} + M_2},
\end{equation}
where $M_{\rm NS}$ and $M_2$ are the NS and giant star masses. The newborn NS through AIC would have a very
small kick (Hurley et al.2010; Tauris et al.2013), thus I do not include the effect of the kick.
The processes of mass accretion and mass loss for the NS binaries are also not well-known (these processes are
out of the main propose of this work; e.g., L$\ddot{\rm u}$ et al. 2012), thus I just
simply assumed that the NS accretes half of the transferred matter, and the rest of
the transferred matter is assumed to be expelled.
Corresponding to the accretion onto the NS ($\dot{M}$ is the accretion rate),
the apparent X-ray luminosity can be calculated as $L_{\rm acc} = \eta \dot{M} c$ (Shakura \& Sunyaev 1973), where $\eta$ is the radiative efficiency
 of accretion onto the NS, and $c$ is the speed of light. The typical luminosity of NS X-ray binaries is $\gtrsim 10^{32}$ erg $\rm s^{-1}$.
 The angular momentum evolution ($\dot{J}$) in the post-AIC evolution
 is mainly dominated by the angular momentum loss due
to mass loss, and is also affected by the gravitational wave radiation and magnetic braking,  (and $\dot{J}_{\rm ML}$, $\dot{J}_{\rm GR}$  and $\dot{J}_{\rm MB}$ ), which is
$\dot{J} =  \dot{J}_{\rm GR} + \dot{J}_{\rm ML}$ + $\dot{J}_{\rm MB}$.

Figure 4 shows the properties of the binaries and the donor stars at the time just after the AIC.
From the AIC with highly magnetized WDs, newborn NSs in these binaries are strongly magnetized (even magnetars) as
discussed in the above sections, and they can be symbiotic X-ray sources
after the highly magnetized NSs start mass accretion via stellar
wind from the giant star.
The ranges of companion masses and orbital periods for the SyXBs with
these newly born highly magnetized NSs (dashed line area in Figure 4)
are 5-652 days and 0.41-1.21 $M_\odot$, respectively.
For the SyXBs formed with the non-magnetic NSs (solid line area),
the ranges of companion masses and orbital periods
are 21.3-2521.5 days and 0.59-1.23 $M_\odot$, respectively. There is a partially overlapped region in Figure 4, I suggest that finding lower-mass and relatively more luminous giant stars after the AIC may provide important clue to understand the magnetic case of this symbiotic channel. Besides,  companions of newborn NSs after the AIC will evolve to He WDs, therefore finding low mass He WDs - NS systems may also be crucial for understanding this symbiotic progenitor scenario.
Observational results for the binary and donor properties of
4U 1700+24 and IGR J17329-2731 are not available yet,
and only the orbital period of 4U 1700+24 (404 days; Masetti et al. 2002) is known.
Our results from AIC under the non-magnetic and magnetic cases
can reproduce the orbital period of 4U 1700+24 (dotted line in Figure 4).
I expect that future observations will provide
more parameters of these known SyXBs to test the AIC models in this work.

\subsection{Discussion}

Compared to previous studies (e.g. Ivanova et al. 2008; Sutantyo \& Li 2000), our model in this work has following advantages
to explain the existence of highly magnetized NSs in old stellar populations like 4U 1700+24 and IGR J17329-2731:
1. The presence of a strong magnetic field and its crucial role in the pre-AIC
evolution is more likely to help for producing highly magnetized NSs in SyXBs.
The conservation of B-field might be uncertain, but what would take
away or decay that high magnetic flux during the AIC process? The mass
accretion in the pre-AIC or the mass loss during the AIC is not enough to decrease so magnetic flux stays high.
2. The timescale of the pre-AIC evolution in this work is very long (9-10 Gyr), thus the highly magnetized NSs
can be born in old systems in wider orbits.
3. The newborn NSs also can be low rotators to avoid the propeller effect, so that the newly born and highly magnetized NSs can
shine as X-ray sources through the wind mass accretion from the giant companions, and this does
not need a long timescale (no too much accretion) and cannot decay the magnetic field of NSs significantly.

These X-ray binaries with highly magnetized NSs formed via AIC may evolve to become ultra-luminous X-ray sources (e.g. Abdusalam et al. 2020)
In addition, the AIC of a highly magnetized WD also may form fast radio burst
sources with a young and highly magnetized NS such as FRB 20200120E (Kirsten et al. 2021).
It has also been studied that the AIC channel can form magnetars in binaries (Ablimit 2022).
Recent observational results for CXOU J171405.7-381031 and SGR 0755-2933 (Chrimes et al. 2022) show that these
two candidates may have stellar companions associated with magnetars, and one
alternative evolutionary pathway to form this kind of candidate may be the AIC channel.

The AIC channel suffers from two main disadvantages (e.g. Kitaura et al. 2006; Dessart et al. 2006):
(1) The lack of direct observational evidences for AIC. (2) The uncertainties/unclearness about the spin rate
and the magnetic field strength of a newborn NS formed via AIC.  The rotational velocities and the magnetic activity depend on
the binary evolutionary history (Soker 2002).
Tauris \& van den Heuvel (2006) demonstrated that a Chandrasekhar mass WD with an equatorial radius of about
3000 km and a magnetic field strength of $10^3$ G could undergo AIC and
directly produce an NS with B $10^8$ G by assuming flux conservation.
Thus, producing a strongly magnetized NS (at the level of $10^{12}$ G) through
the AIC of a highly magnetized WD (as used in this work) is not a problem.
Besides, Dessart et al. (2006) studied that only
a few 0.001 $M_\odot$ of material is ejected during the AIC, thus this small amount of
mass ejection indicates the higher possibility of conservation.
However, it is also the reason why the AIC is hard to be observed (see Dessart et al. (2006) and Piro \& Kulkarni (2013) for more discussion).
The another uncertainty is the accretion rate and magnetic field decay in the post-AIC evolution.
It is still not well-known that the newborn NS could accrete how many percent of the transferred material,
and how the accretion, magnetic field and spin (propeller effect) will affect the evolution.
Compared to other formation pathways, to our current knowledge, the SyXBs with strongly magnetized
NSs is more likely formed via the AIC channel.

\section{Conclusion}

Recent observational results show that IGR J17329-2731, one of the known SyXBs,
contains a highly magnetized NS,
and suggest that another SyXB 4U 1700+24 may also have a strongly magnetized NS.
Most previous theoretical formation scenarios fail to explain the presence of strongly magnetized NSs in SyXBs.
The AIC channel from highly magnetized ONeMg WDs -- RG binaries is probably the promising formation channel to produce SyXBs with strongly magnetized NSs.
However, the detailed evolution for this AIC channel has not yet been studied.
In this work, with the \textsc{mesa} stellar evolution code to simulate large grids of stellar and binary evolution,
I have investigated both the pre- and post-AIC binary evolution to show that the AIC channel from WD--RG binaries is the
favorable way to form SyXBs with strongly magnetized NSs.
In the pre-AIC evolution of ONeMg WD - red giant star binaries, I consider cases where the ONeMg WD is
non-magnetic and cases where the magnetic field is strong and forces accretion onto
two small caps on the poles of the WD. Performing stellar evolution simulations our findings are as follows:

\begin{itemize}
\item  If the ONeMg WD in the AIC channel is strongly magnetized, the magnetism has a crucial role in the pre-AIC evolution. In the AIC evolution, the newborn NS may inherit the large magnetic field through conservation of magnetic flux, and the newly formed highly magnetized NS system has a long age like the RG companion (9-11 Gyr). The wind accretion of the NS (without significant magnetic field decay) from the old giant companion can shine as an SyXB which is similar to the observations of IGR J17329-2731.
\item  When the ONeMg WD that accretes mass
from an RG star has a strong magnetic field, lower mass RG stars can lead to AIC, because the confined accretion onto a small area allows
more efficient hydrogen burning to suppress nova outbursts. As a result of the
magnetic confinement, the WD mass can grow to the Chandrasekhar mass limit and  collapse as showed in this work or explode
as an SN Ia in the symbiotic channel of the single degenerate scenario (Ablimit et al. 2023). Our results
also show that, with the magnetic confinement the initial parameter space of this
symbiotic channel for SNe Ia shrinks toward shorter orbital periods ($5-120$ days) and lower donor
masses ($0.7-1.8\,M_\odot$) compared with that ($10-750$ days and $1.1-1.8\,M_\odot$) in the no magnetic confinement case.
\item After the AIC evolved from the symbiotic channel with different cases, newborn NSs can be normal NSs or peculiar NS objects as pulsars or magnetars. 
For observations, searches for relatively lower-massive giant stars or He WDs as companions of these NSs will be helpful to improve our understanding on this scenario.
\end{itemize}

\section*{Acknowledgements}

I thank the referee for the useful comments that improved this work.
This work is supported by NSFs.

\section*{Data Availability} The data underlying this article
will be shared on reasonable request to the corresponding author.





\clearpage

\textbf{REFERENCES}

Abdusalam, K., Ablimit, I., Hashim, P., L$\ddot{\rm u}$, G.-L. et al. 2020, ApJ, 902, 125

Ablimit, I. \& Li, X.-D., 2015, ApJ, 800, 98

Ablimit, I., \& Maeda, K. 2019a, ApJ, 871, 31

Ablimit, I., \& Maeda, K. 2019b, ApJ, 885, 99

Ablimit, I. 2019, ApJ, 881, 72

Ablimit, I. 2022, MNRAS, 509, 6061

Ablimit, I., Podsiadlowski, Ph., Hirai, R. \& Wicker, J.  2022, MNRAS, 513, 4802

Ablimit, I., et al.  2023, submitted 

Angel, J. R. P., Borra, E. F., \& Landstreet, J. D. 1981, ApJS, 45, 457

Belloni, D., Schreiber, M. R., Salaris, M., Maccarone, T. J. \& Zorotovic, M. 2021, MNRAS, 505, L74

Bhattacharya, D., \& van den Heuvel, E. P. J. 1991, PhR, 203, 1

Bozzo, E., Pavan, L., Ferrigno, C., Falanga, M., Campana, S., Paltani, S., Stella,
L., Walter, R., 2012, A\&A, 544, A118

Bozzo, E., Romano, P., Ferrigno, C., Esposito, P., Mangano, V., 2013,
Advances in Space Research, 51, 1593

Bozzo, E., et al., 2018, A\&A, 613, A22

Bozzo, E., Romano, P. Ferrigno, C. \& Oskinova, L. 2022, arXiv:2203.16347

Brooks, J., Schwab, J., Bildsten, L., Quataert, E.,  \& Paxton, B. 2017, ApJ, 843,
151

Camisassa, M. E., Raddi, R., Althaus, L. G. et al. 2022, MNRAS, 516, L1

Chanmugam, G., \& Brecher, K. 1987, Natur, 329, 696

Corbet, R. H. D., Sokoloski, J. L., Mukai, K., Markwardt, C. B., Tueller, J.,
2008, ApJ, 675, 1424

Cropper, M. 1990, SSRv, 54, 195

Dessart, L., Burrows, A., Ott, C. D., et al. 2006, ApJ, 644, 1063

Duncan, R. C., \& Thompson, C. 1992, ApJL, 392, L9

Eggleton, P. P. 1983, ApJ, 268, 368

Fabian, A. C., Pringle, J. E., Rees, M. J. \& Whelan, J. A. J. 1977, MNRAS, 179, 9

Ferrario, L., de Martino, D., \& Gaensicke, B. T. 2015, SSRv, 191, 111F

Frank, J., King, A., \& Raine, D. J. 2002, Accretion Power
in Astrophysics, by Juhan Frank and Andrew King and
Derek Raine, pp. 398. ISBN 0521620538. Cambridge,
UK: Cambridge University Press, February 2002., 398

Freire, P.C.C. \& Tauris, T.M. 2014, MNRAS 438(1), L86

Garcia-Berro, E., Loren-Aguilar, P., Aznar-Siguan, G., et al. 2012, ApJ, 749, 25

Ginzburg, S., Fuller, J., Kawka, A. \& Caiazzo, I. 2022, MNRAS, 514, 4111

Hameury, J.-M., King, A. R., \& Lasota, J.-P. 1986,
MNRAS, 218, 695

Hillman, Y., Prialnik, D., Kovetz, A., \& Shara, M. M. 2015, MNRAS, 446, 1924

Hillman, Y., Prialnik, D., Kovetz, A., \& Shara, M. M. 2016, ApJ, 819, 168

Hurley, J. R., Tout, C. A. \& Pols, O. R. 2002, MNRAS, 329, 89

Hurley, J. R., Tout, C. A., Wichramasinghe, D. T., Ferrario, L., \& Kiel, P. D.
2010, MNRAS, 402, 1437

Isern, J., Garcia-Berro, E., Kvlebi, B. \& Loren-Aguilar, P. 2017, ApJ, 836, L28

Ivanova, N., Heinke, C. O., Rasio, F. A., Belczynski, K., \& Fregeau, J. M. 2008,
MNRAS, 386, 553

Kaplan, D. L., Levine, A. M., Chakrabarty, D., Morgan, E. H., Erb D. K.,
Gaensler, B. M., Moon, D.-S., Cameron, P. B., 2007, ApJ, 661, 437

Kato, M., \& Hachisu, I. 2004, ApJL, 613, L129

Kahabka, P. 1995, ASP Conference Series, Vol. 85

King, A. R., \& Ritter, H. 1999, MNRAS, 309, 253

King, A. R. \& Shaviv, G. 1984, MNRAS, 211, 883

Kitaura, F. S., Janka, H.-T., \& Hillebrandt, W. 2006, A\&A, 450, 345

Kolb, U., Davies, M. B., King, A. \& Ritter, H. 2000, MNRAS, 317, 438

Kulkarni, S. R., \& Narayan, R. 1988, ApJ, 335, 755

Levy, E. H., \& Rose, W. K. 1974, ApJ, 193, 419

Livio, M. 1983, A\&A, 121, L7

L$\ddot{\rm u}$, G., Zhu, C.,Postnov, L. R. et al. 2012, MNRAS, 424, 2265

Masetti, N. et al. 2002, A\&A, 382, 104

Masetti, N., Orlandini, M., Palazzi, E., Amati, L., \& Frontera, F. 2006, A\&A,
453, 295

Michel, F. C. 1987, Natur, 329, 310

Mukhopadhyay, B. \& Bhattacharya, M. 2022, Particles 5, 493

Nauenberg, M. 1972, ApJ, 175, 417

Nespoli, E., Fabregat, J., Mennickent, R. E., 2008, A\&A, 486, 911

Nomoto K. \& Iben I., 1985, ApJ , 297, 531

Nomoto, K., Miyaji, S., Sugimoto, D., \& Yokoi, K. 1979, in White Dwarfs and
Variable Degenerate Stars, eds. H.M. van Horn, \& V.Weidemann, IAU Coll.,
53, 56

Nomoto, K., Saio, H., Kato, M., \& Hachisu, I. 2007,ApJ,663, 1269

Nordhaus, J., Wellons, S., Spiegel, D. S., Metzger, B. D., \& Blackman, E. G.
2011, PNAS, 108, 3135

Norton, A.J., Watson, M.G. 1989, MNRAS, 237, 715

Osborne, et al. 2001, A\&A, 378, 800

Paxton, B., Bildsten, L., Dotter, A., et al. 2011, ApJS, 192, 3

Paxton, B., Marchant, P., Schwab, J., et al. 2015, ApJS, 220, 15

Perets, H. B. 2022, ApJ, 927, L23

Podsiadlowski, Ph., \& Rappaport, S. 2000, ApJ, 529, 946

Piro, A. L., \& Kulkarni, S. R. 2013, ApJ, 762, L17

Sabach, E. \& Soker, N. 2014, MNRAS, 439, 954

Schreiber, M. R.,  Belloni, D.,  Gansicke, B. T., Parsons, S. G. \& Zorotovic, M. 2021, Nature Astronomy, 5, 648

Shakura, N. I. \& Sunyaev, R. A. 1973, A\&A, 500, 33

Schwab, J., Podsiadlowski, P., \& Rappaport, S. 2010, ApJ, 719, 722

Schwope, A., Buckley, D. A.H., Malyali, A. et al. 2021, eprint arXiv:2106.14540

Shibazaki, N., Murakami, T., Shaham, J., \& Nomoto, K. 1989, Nature, 342, 656

Soker, N., 2002, MNRAS, 337, 1038

Sokoloski, J. L., \& Beldstin, L. 1999, ApJ, 517, 919

Sutantyo, W., \& Li, X.-D. 2000, A\&A, 360, 633

Taam, R. E., \& van den Heuvel, E. P. J. 1986, ApJ, 305, 235

Tauris, T. M., Sanyal, D., Yoon, S.-C., \& Langer, N. 2013, A\&A, 558, A39

Tauris, T. M., \& Savonije, G. J. 1999, A\&A, 350, 928

Tauris, T. M., \& van den Heuvel, E. P. J. 2006, in Formation and evolution of
compact stellar X-ray sources (Cambridge University Press), 623

Tout, C. A., Wickramasinghe, D. T., Liebert, J., Ferrario, L., \& Pringle, J. E.
2008, MNRAS, 387, 897

van den Heuvel, E. P. J. 1994, in Interacting Binaries, eds. H. Nussbaumer, \&
A. Orr (Berlin: Springer), 263

Verbunt, F., Wijers, R. A. M. J., \& Burn, H. M. G. 1990, A\&A, 234, 195

Wang, B. \& Liu, D. 2020, Research in Astronomy and Astrophysics, 20, 135

Wang, B., Liu, D. \& Chen, H.-L. 2022, MNRAS, 510, 6011

Wickramasinghe, D. T. \& Ferrario, L. 2000, PASP, 112, 873

Wickramasinghe, D. 2014, European Physical Journal Web of Conferences, 64,
03001

Wolf, W. M., Bildsten, L., Brooks, J. \& Paxton, B. 2013, ApJ, 777, 136

Yoon S.-C., Langer N., 2005, A\&A, 435, 967


\clearpage

\begin{figure*}
\centering
\includegraphics[totalheight=4.7in,width=3.8in]{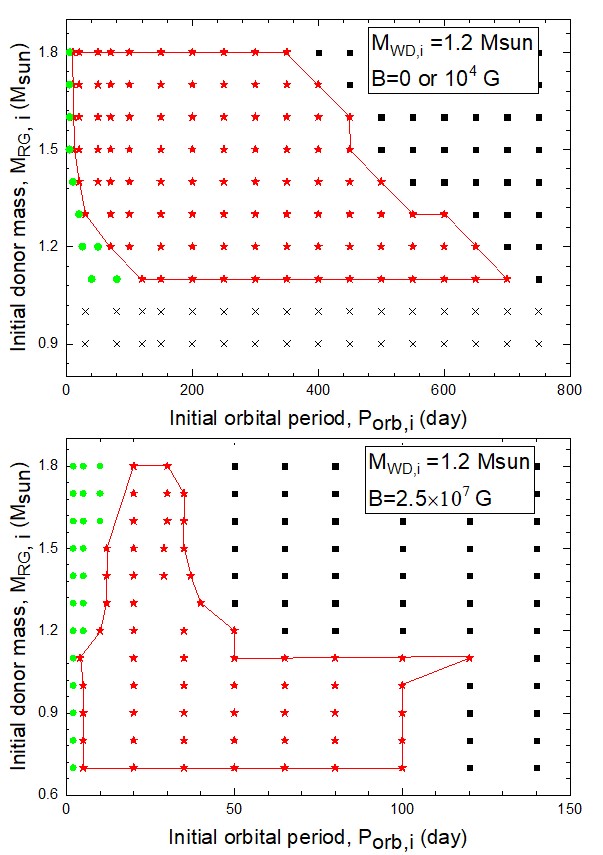}
\caption{Initial parameter space of ONeMg WD $+$ RG star binaries which lead to
AIC in the initial orbital period ($P_{\rm orb,i}$)-initial secondary mass ($M_{\rm He,i}$) plane. The upper panel shows results under the magnetic confinement case, and the lower panel  is  for the no magnetic confinement case (non- or low magnetic WDs). The initial WD masses are 1.2 $M_\odot$ for both cases.
Red stars represents for binaries which can lead to AIC, while other symbols show that these WD + RG binaries cannot produce AIC.
The regions encircled by solid red lines are for producing AIC from WD + RG systems.
The upper limits of the regions are based on the occurrence of a common
envelope. The lower limits (i.e. black cross) are determined by the
initiation of nova bursts. The left and right limits (i.e. green circles and black square) are derived
according to inefficient mass transfer.}
\end{figure*}

\clearpage

\begin{figure*}
\centering
\includegraphics[totalheight=5.7in,width=5.3in]{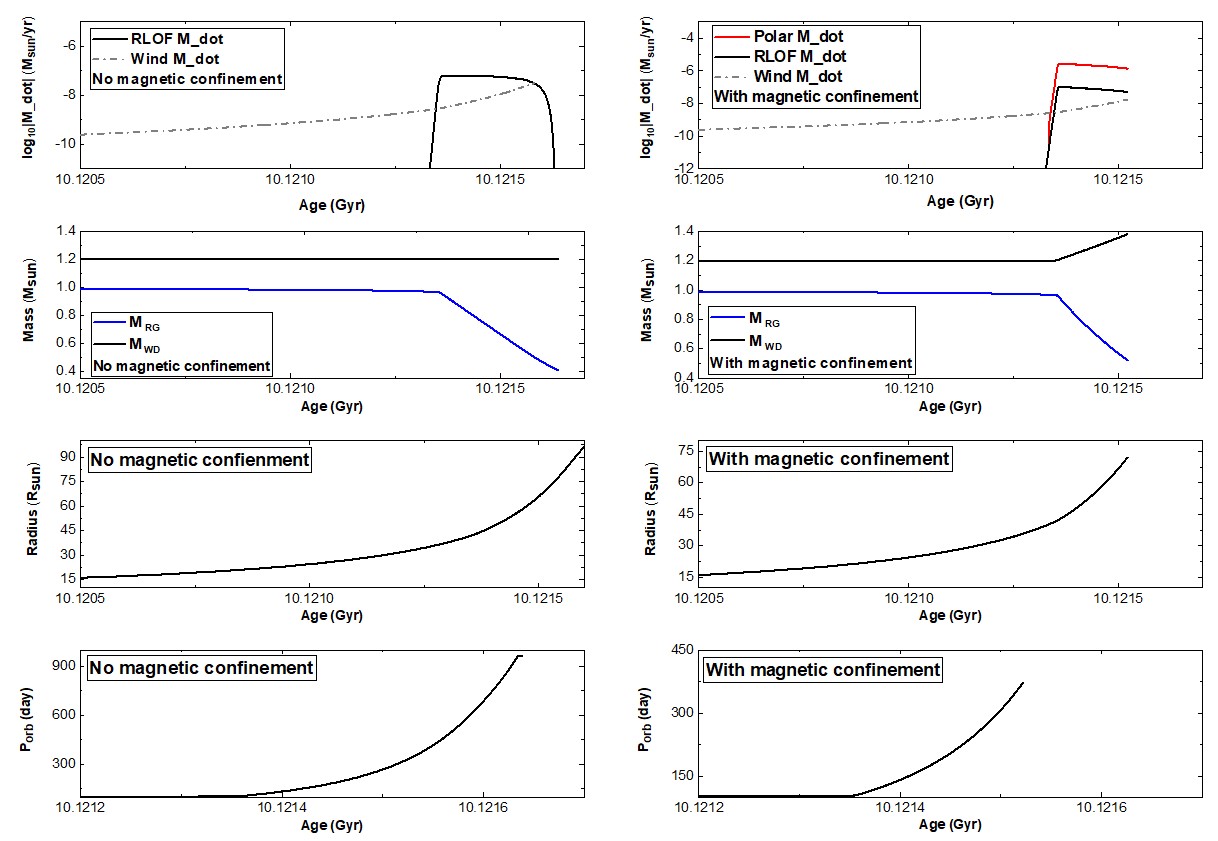}
\caption{Pre-AIC evolutions of ONeMg WD $+$ RG binaries: the left panels for the no magnetic
confinement case, and right panels are for the magnetic confinement case (a highly magnetized WD with B$=2.5\times10^7$ G.
The initial masses of the WD and RG donor star are 1.2 $M_\odot$ and 1.0 $M_\odot$ in this example, respectively.
The initial orbital period is 100 days. Evolutions of mass transfer including the wind mass loss (polar mass accretion for the magnetic confinement case), the WD/donor mass, orbital period, and the radius evolutions of the donors with time are shown in the figure.
The wind mass loss is low ($<1\times10^{-8} M_\odot/{\rm yr})$ almost in the whole evolution time, and RLOF mass transfer dominates the evolution.}
\end{figure*}

\clearpage

\begin{figure*}
\centering
\includegraphics[totalheight=5.7in,width=5.3in]{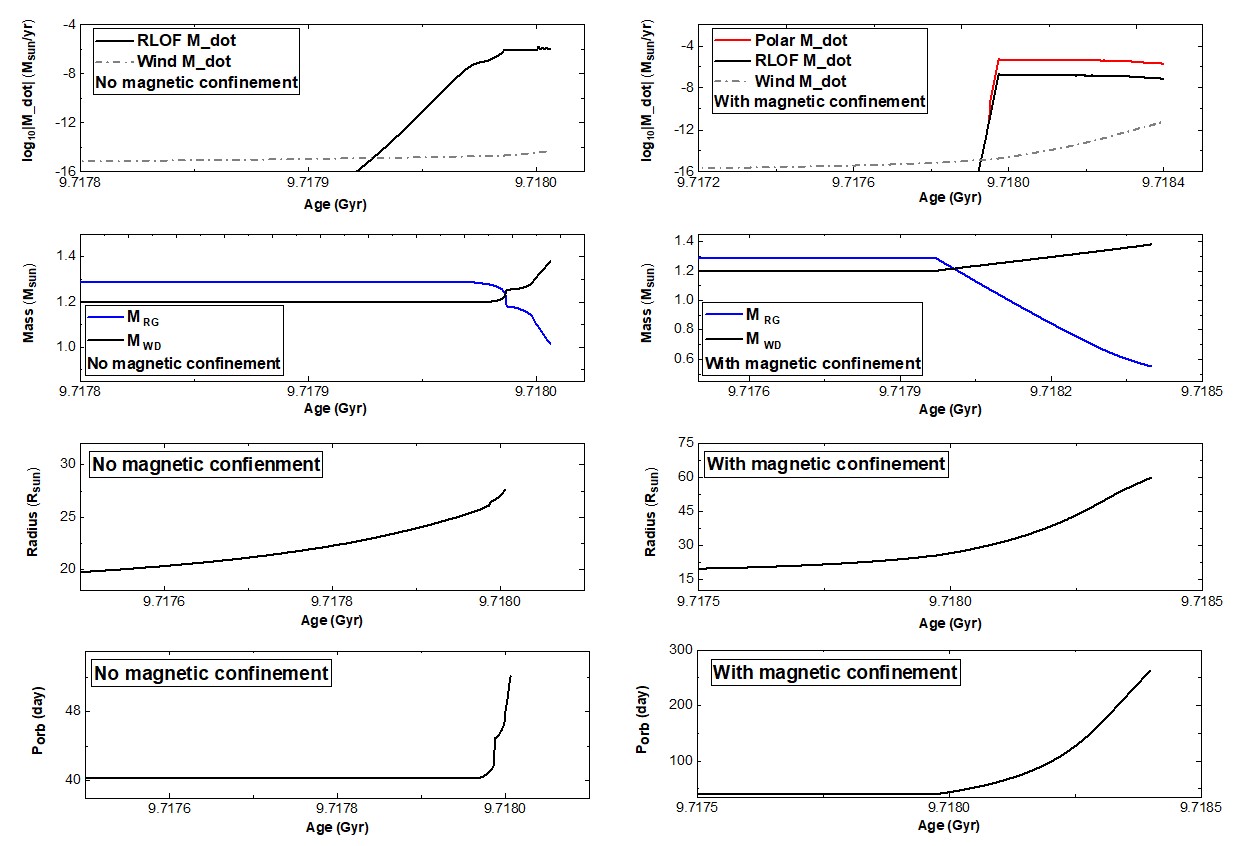}
\caption{Another example for detailed evolution of ONeMg WD $+$ RG star binaries using the MESA code:
evolutions of mass transfer (polar mass accretion for the magnetic case), wind mass loss,
the WD/donor mass, orbital period , and the radius evolutions of the donors  with time are given. Right and left panels show
the non-magnetized (no magnetic confinement case) and highly magnetized (B$=2.5\times10^7$ G; with magnetic confinement case) WD cases, respectively. The
initial masses of the WD and RG donor are 1.2 $M_\odot$ and 1.3 $M_\odot$, and the initial orbital period is 40 days.
The wind mass loss is very low ($<10^{-11} M_\odot/{\rm yr})$ which only affects the angular momentum evolution in a relatively significant way.
}
\end{figure*}

\clearpage

\begin{figure*}
\centering
\includegraphics[totalheight=4.0in,width=4.6in]{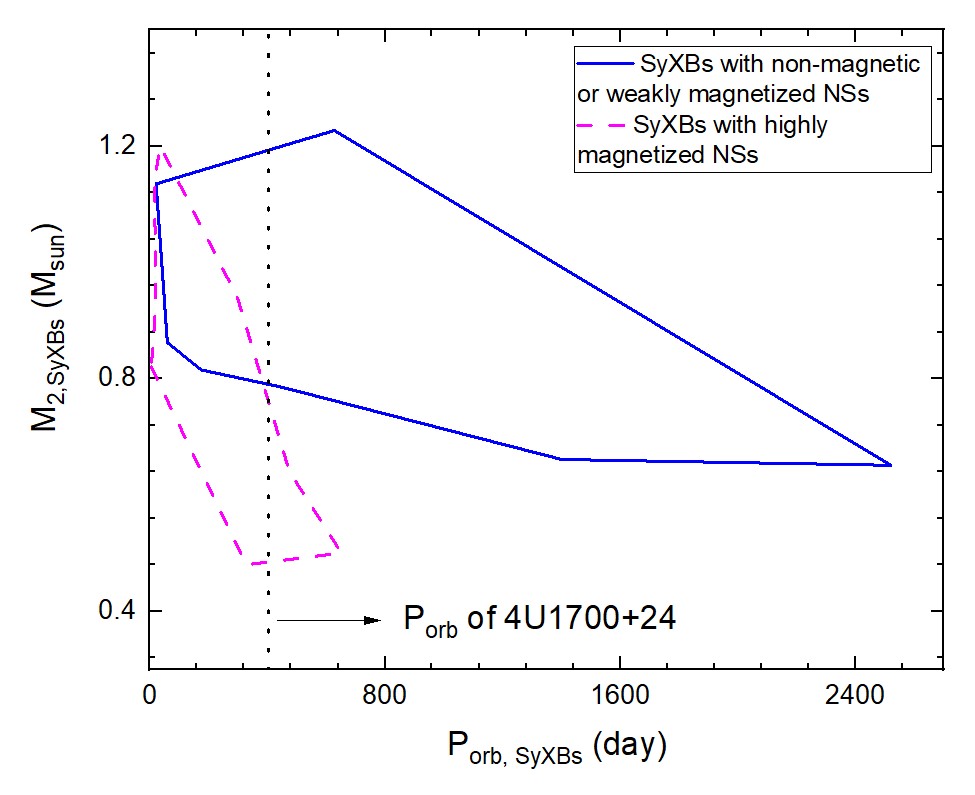}
\caption{Companion giant star mass distributions versus Orbital periods for
SyXBs evolved from ONeMg WD $+$ RG binaries of Figure 1.}
\end{figure*}

\clearpage







\label{lastpage}
\end{document}